\documentclass[conference]{IEEEtran}

\usepackage[utf8]{inputenc}
\usepackage[backend=biber,style=ieee,url=false,isbn=false]{biblatex}
\usepackage{amsfonts}
\usepackage{amsthm}
\newtheorem{lemm}{Lemma}
\usepackage{algorithmic}
\usepackage{graphicx}
\usepackage{textcomp}
\usepackage{xcolor}
\usepackage{listings}
\usepackage{todonotes}
\usepackage{subcaption}
\usepackage{float}

\begin{document}

\title{Linearizability and State-Machine Replication: \\Is it a match?\\\small{~}\\\small{Version of \today}}

\author{\IEEEauthorblockN{Franz J. Hauck}
   \IEEEauthorblockA{\textit{Institute of Distributed Systems} \\
   \textit{Ulm University}\\
   0000-0002-7480-9617\\
   }
\and
   \IEEEauthorblockN{Alexander Heß}
   \IEEEauthorblockA{\textit{Institute of Distributed Systems} \\
   \textit{Ulm University}\\
   0000-0001-6837-2861\\
   }
}

\maketitle
\thispagestyle{plain}
\pagestyle{plain}

\begin{abstract}
Linearizability is a well-known correctness property for concurrent and distributed systems.
In the past, it was also used to prove the design and implementation of replicated state-machines correct.
State-machine replication (SMR) is a concept to achieve fault-tolerant services by replicating the application code and maintaining its deterministic execution in all correct replicas.
Correctness of SMR needs to address both, the execution of the application code---the state machine---and the replication framework that takes care of communication, checkpointing and recovery.
We show that linearizability is overly restrictive for SMR as it excludes many applications from being replicated.
It cannot deal with conditional waits and bidirectional data flows between concurrent executions.
Further, linearizability does not allow for vertical composition, e.g., nested invocations.
In this work we argue that interval linearizability, a recently defined relaxation of linearizability, is the appropriate correctness criterion for SMR systems.
This no longer puts any constraints on the application code.
Instead the focus for correctness proofs of an SMR system is on the deterministic execution of the application within correct replicas under the assumptions of the chosen failure model.
Thus, this work not only clears some myths about linearizable SMR but also relieves correctness proofs from linearizability proofs.
\end{abstract}

  
\section{Introduction}

Linearizability is a well-known and popular correctness criterion for concurrent and distributed systems~\cite{herlihy_linearizability_1990,sela_brief_2021}.
It binds atomic request execution to real-time, and as a consequence the effect of a given request execution has to be observable by subsequent invocations.
For distributed systems, linearizability guarantees that the system behaves like a non-distributed but likely concurrent system.
This makes linearizability a good candidate for proving distributed systems correct.

State-machine replication (SMR) was first sketched by Lamport~\cite{lamport_time_1978}, but the tutorial written by Schneider~\cite{schneider_implementing_1990} is probably the most cited fundamental work.
SMR achieves fault tolerance by replicating a service on multiple nodes in a distributed system.
Each replica receives the same client requests---typically in the same total order---and executes them deterministically.
This ensures that all correct replicas have eventually the same state, and can provide a correct answer to clients.
SMR is used to build fault-tolerant systems based on the crash-stop or crash-recovery failure model (CFT), but is even suitable for the Byzantine failure model (BFT) that allows for almost any faulty behaviour in a defined subset of the total number of replicas.

Linearizability was extensively used in the past to prove SMR systems correct.
For example, the seminal work about the first practical implementation for BFT SMR by Castro, called PBFT, takes linearizability as the correctness property of choice~\cite{castro_practical_2001}.

In this paper, we argue that linearizability is too strict to be a correctness criterion for SMR systems.
Services that deploy fine-grained locking to protect their shared data structures or need to use conditional waits to coordinate internal resources and block their threads are no longer linearizable.
Likewise, nested invocations, i.e. replicated services that call other replicated or non-replicated services, are not linearizable in the vast majority of cases.
Further, systems with concurrent threads that communicate bi-directionaly are not linearizable.
Although linearizability is a match for data-oriented systems that basically comprise read and write operations, it is not for many others.
Next to discussing the shortcomings of linearizability for SMR, we will argue that interval linearizability~\cite{castaneda_unifying_2018} is \emph{the} appropriate correctness criterion for SMR, although it might be interesting to stick to linearizability when it comes to particular SMR applications, e.g. storage systems that are supposed to have atomic read and write operations.

Linearizability of SMR systems cannot be proven without assumptions about the replicated application.
We show that these assumptions vanish as soon as interval linearizability is the criterion of choice.
Instead, proofs of correctness can focus on the replication framework and exclude the application or service code.
In principle, the framework has to ensure deterministic execution of the application in all correct replicas under the assumption of the chosen failure model.

In Section~\ref{sec:background}, we provide informal definitions of linearizability, interval linearizability and SMR.
Section~\ref{sec:history} describes the application of linearizability to SMR systems in a historical context.
This section also collects a couple of false or at least misleading statements about linearizability and SMR, made in scientific papers partially published in highly ranked conferences.
In Section~\ref{sec:match}, we critically look at linearizability and its application to SMR systems.
In Section~\ref{sec:solution}, we will suggest interval linearizability~\cite{castaneda_unifying_2018} as the appropriate correctness criterion and show that it can cover all those SMR applications that linearizability cannot.
In Section~\ref{sec:proof}, we describe the consequences for proving SMR systems correct.
Finally, Section~\ref{sec:conclusion} provides our conclusions.

\section{Background}
\label{sec:background}

First, we will recap the definition and meaning of linearizability for concurrent and especially distributed systems.
Second, we will introduce interval linearizability as a very generic relaxation of classic linearizability.
Finally, state-machine replication is explained.

\subsection{Linearizability}

The concept of Linearizability was introduced by Herlihy and Wing in 1987~\cite{herlihy_axioms_1987} and later refined in 1990~\cite{herlihy_linearizability_1990}.
As the definition of linearizability contained an undetected typo that rendered the definition wrong, at least for some corner cases, the typo was fixed and reported in 2021~\cite{sela_brief_2021}.

The system model for linearizability is that there is an object, offering operations which can be invoked concurrently by an arbitrary number of clients and may be even concurrently executed within the object implementation.
The object may contain shared state that can be accessed only by the operation executions.
This is in line with classic object-based programming models.
For the definition of linearizability, it is important to consider the invocation event, where a client invokes the object operation, and the response event, where a client receives the corresponding response from the object.

Linearizability now provides a guarantee about the expectable behaviour of the operation executions.
More precisely and without repeating a formal definition which can be found in~\cite{sela_brief_2021}, linearizability requires the effect of an invocation to be equivalent to an atomic event that happens between the invocation event and the response event, i.e. all effects of the operation happen all at once at this atomic event, sometimes called a \emph{linearization point}.\footnote{The concept of linearization points seems to have evolved from \emph{atomic objects}~\cite{lynch_distributed_1996}.}
Further, the linearization points of all concurrent invocations are totally ordered in time, i.e. never happen concurrently nor at the same time.

Note that in general it is practically impossible to compress all effects of an operation into a single atomic event.
However, linearizability requires the equivalence of the actual behaviour, i.e. the object implementation may even allow concurrent executions and concurrent accesses to shared state as long as the observable behaviour is equivalent to a sequence of atomic linearization points, one per invocation.

\begin{figure}[ht]
  \center
  \begin{subfigure}{.4\textwidth}
    \center
    \includegraphics[width=.7\textwidth]{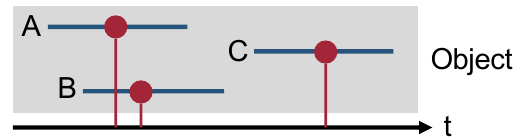}
    \caption{Sequence of linearization points}
    \label{fig:lin.1}
  \end{subfigure}\\
  \vspace{0.3cm}
  \begin{subfigure}{0.4\textwidth}
    \center
    \includegraphics[width=.7\textwidth]{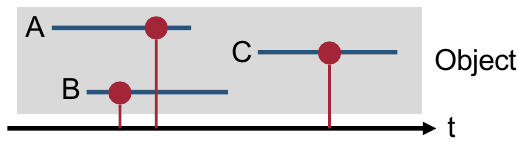}
    \caption{Alternate sequence}
    \label{fig:lin.2}
  \end{subfigure}
  \caption{Three invocations with their linearization points}
  \label{fig:lin}
\end{figure}

Figure~\ref{fig:lin.1} shows the timeline of three invocations A, B and C including their invocation and response event, drawn as the start and end of a line representing the execution of the corresponding operations.
On each of these lines there is the conceptual linearization point marked as a red circle.
Since invocations A and B overlap, their linearization points can be arranged in two different orders (cf. Figure~\ref{fig:lin.2}).
As invocation C does not overlap with neither A nor B and as it is processed after both invocations are completed, linearizability guarantees that the corresponding linearization points are arranged in exactly that order.

In the following, we will discuss the consequences and effects of a linearizable system.

\subsubsection{Consistency}

If the object is a memory cell, memory bank or some sort of storage system that basically offers read and write operations, linearizability renders the system to be \emph{strictly consistent}.
That is why linearizability is sometimes considered to be equivalent to strict consistency.
With strict consistency, read operations read the value stored by the most recent write operation before the read.
It is obvious that a linearizable object will guarantee exactly that.
If invocations A and B are write operations with different values, and invocation C is a read operation, C will read the value of the most recent write depending on the actual order of their linearization points.

\subsubsection{Serializability}

Linearizability has similar properties as \emph{serializability}, a well-known correctness criterion of transactional systems (cf.~\cite{papadimitriou_serializability_1979}).
Both require that the effects of operations, or transactions respectively, have to be atomic and in a timely sequence.
However, linearizability puts constraints on the timely occurrencies of linearization points as they can only occur between the corresponding invocation and response events whereas serializability does not necessarily require that.
An extension called \emph{strict serializability}~\cite{herlihy_linearizability_1990} at least requires that a transaction has to see the effects of previously completed transactions, which makes it more comparable to linearizability.
Nevertheless, serializability is defined for a transaction containing many computations including operation invocations, whereas linearizability is defined for the individual operations within a particular object.
Linearizability means that some concurrent invocations in an object are equivalent to their sequential execution.
The specification of the object can be sequential whereas the implementation may allow some internal concurrency.


\subsubsection{Distributed Systems}

Whenever the object has a distributed implementation, linearizability guarantees that its behaviour is equivalent to a non-distributed object.
This is not automatically the case.
As an example, a replicated quorum-based storage system offering read and write operations, e.g. the one described in~\cite{merideth_selected_2010} is not linearizable without further efforts.
A read invocation has to contact a quorum of replicas and returns the most recent of the values provided by them.
A write invocation has to store the value at least in a quorum of replicas with the current timestamp.
Such a system is not linearizable due to effects of delayed messages at the communication to (some of) the replicas.
Some client may read a newly written value and after that another client may read an old value depending on the quorum sets they are using.
However, the system can be made linearizable by some measures called \emph{read repair}, also described in~\cite{merideth_selected_2010}.
With read repair, a read operation has to immediately write the read value back to a quorum of nodes within its execution.
This ensures that after the read value is delivered to the caller, no older value can be read.

In summary, linearizability became a popular criterion for services that are implemented by distributed components but are supposed to behave like a non-distributed service.



\begin{figure}[ht]
  \center
  \includegraphics[width=.31\textwidth]{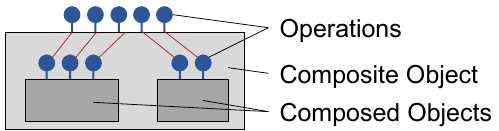}
  \caption{Horizontal composition of two objects}
  \label{fig.horizontal}
\end{figure}

\subsubsection{Local Property}

Linearizability is proven to have a local property~\cite{herlihy_linearizability_1990}.
This allows for composition of multiple linearizable objects, and the composite object will be linearizable, too, without further proof.
The composite object provides the union of operations of all contributing objects.
We will call this horizontal composition, as the composed objects are horizontally contributing to the composite (cf. Fig.~\ref{fig.horizontal}).

\subsection{Interval Linearizablity}

There is quite some work that contributed to relaxed variants of linearizability.
We will concentrate on \emph{interval linearizability} which was introduced by Castañeda et al.~\cite{castaneda_unifying_2018} as a superset of \emph{set linearizabilty}.
Set linearizabilty is in turn a superset of linearizability (linearizability hierarchy):~~\emph{Linearizability} $\subset$ \emph{Set Linearizability} $\subset$ \emph{Interval Linearizability}

\begin{figure}[ht]
  \center
  \includegraphics[width=.3\textwidth]{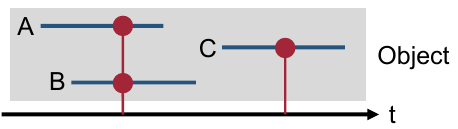}
  \caption{Set-linearizable execution example}
  \label{fig:setlin}
\end{figure}

In all cases, the system model remains the same.
The guarantee of set linearizability is relaxed in comparison to classic linearizability in the sense that linearization points of different executions might happen at the same time, forming a set of linearization points at some point in time.
Within this set, arbitrary data exchange between the corresponding executions is possible.
Set linearizability was informally introduced by Neiger~\cite{neiger_set-linearizability_1994} and formally defined by \cite{castaneda_unifying_2018}.
Fig.~\ref{fig:setlin} shows an example of a set-linearizable execution.

\begin{figure}[ht]
  \center
  \includegraphics[width=.3\textwidth]{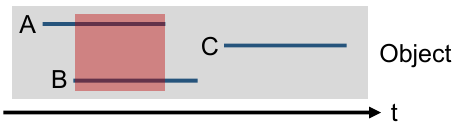}
  \caption{Interval linearizability with highlighted overlap interval}
  \label{fig:inlin}
\end{figure}

The guarantee provided by interval linearizability is even more relaxed compared to set linearizability.
Without repeating the lengthy formal definition, which can be found in~\cite{castaneda_unifying_2018}, interval linearizability allows overlapping invocations, e.g. A and B in Figure~\ref{fig:inlin}, to exchange arbitrary information via shared object state during the time interval of their overlap, highlighted in red.
The effect of a particular invocation is no longer atomic but may happen continuously during the execution interval.

Thus, the specification of an interval-linearizable object is no longer sequential but concurrent.
However, the binding to real time is still there.
Operation C in Figure~\ref{fig:inlin} has to see the concurrently created effects of A and B, as they happened before the invocation of C.
This is especially interesting for distributed systems, as such systems will appear as a single concurrent objects as long as their executions are interval linearizable.

Further, Castañeda et al.~\cite{castaneda_unifying_2018} showed that interval linearizability has still the local property as discussed for classic linearizability.
Even more important is the proof by Goubault et al.~\cite{goubault_concurrent_2018} that \textbf{any concurrent specification for an object is interval linearizable}, which makes this property especially interesting for our context as we will see in Section~\ref{sec:match}.

\subsection{State-Machine Replication}
\label{sec:smr}

As already mentioned, Lamport introduced the idea of a replicated state machine in 1978~\cite{lamport_time_1978}.
Schneider elaborated on the subject in 1990 in his famous tutorial~\cite{schneider_implementing_1990}.
The idea of SMR is to use a deterministic state machine as a represention of a stateful service---in the context of linearizability an object---that offers operations which can be invoked by clients.
The implementations of these state machines can now be replicated; invocations have to be delivered deterministically to all replicas, have to be executed deterministically, and the results have to be returned to the clients by each replica.
Thus, a group of replicas can tolerate a certain number of crash faults or even Byzantine faults.
The latter means that replicas may expose almost any malicious behaviour including collusion between multiple malicious replicas, trials to compromise the deterministic delivery of invocations, and falsifying answers to clients.
However, certain aspects are typically excluded from malicious behaviour, e.g., breaking current cryptography including secure hashes, encryption and signature schemes.
In the Byzantine failure model, a client can no longer rely on a single response, but has to do some voting on the received responses from replicas in order to identify the correct version among the responses.
Typically $f+1$ equal responses mark the correct version as only up to $f$ replicas are supposed to be faulty.

The deterministic input for the state machines is typically enforced by an atomic broadcast based on a consensus protocol.
The consensus protocol has the advantage compared to other approaches that it is already designed for some failure model and can thus tolerate faults within that model.
Further, a consensus protocol has its own assumptions about the distributed system, e.g. asynchrony or partial synchrony, that have to apply.
For crash faults, a consensus protocol can tolerate up to $\lfloor\frac{n-1}{2}\rfloor$ faulty replicas with a total of $n$ replicas.
For Byzantine faults, typically only up to $\lfloor\frac{n-1}{3}\rfloor$ faults can be tolerated.\footnote{There are exception with the help of trusted computing environments, which we leave out here as our results will be equally applicable.}
An SMR system with more faults than allowed by the failure model is outside of its specification and most likely fails.

An atomic broadcast delivers the invocations of all existing clients to all replicas in the same total order, and thus achieves deterministic input to the service object.
The service object is supposed to implement a deterministic state machine that executes the same state transitions on these invocations.
Correct replicas thus have the same state and can deliver the same responses to clients.

Schneider considered the term state machine for the implementation of a service object to be somewhat ``poor"~\cite{schneider_implementing_1990}.
In practice, the service object is implemented in some programming language and mostly not explicitly modelled as a state machine.
In any case, the execution of the application code has to be deterministic, and typically some restrictions have to be applied, e.g., avoidance of non-deterministic instructions, statements, operations and system calls.
Further, the execution in most SMR systems is sequential, i.e. each replica executes one request invocation after the other.

\begin{figure}[h]
  \center
  \includegraphics[scale=.62]{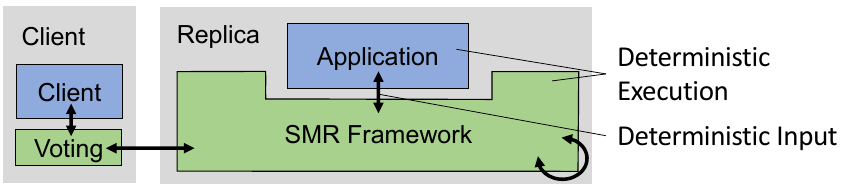}
  \caption{Client and replica component}
  \label{fig.framework}
\end{figure}

In modern SMR systems, however, more advanced concepts are applied.
Instead of an atomic broadcast, some modern egalitarian consensus algorithms do not deliver requests in a total order to replicas but in a partial order (e.g.,~\cite{moraru_there_2013,eischer_egalitarian_2021}).
Unrelated requests can be identified using a-priori knowledge, and as a consequence, their execution order will not effect the state of the service object.
Furthermore, instead of sequential executions deterministic multithreading may be used to achieve deterministic executions.
A plethora of work was created on how to achieve concurrent but deterministic execution of request invocations (e.g.,~\cite{basile_active_2006,reiser_consistent_2006,kapritsos_all_2012,hauck_uds_2016,alchieri_boosting_2018,kapitza_storyboard_2010}).

Without loss of generality, in this paper, we separate an SMR system into the application code, i.e. the service object, and the replication framework, i.e. all components that maintain the replication protocol and the deterministic execution within the service object.
Application and framework code together comprise one of the replicas of an SMR system (see Figure~\ref{fig.framework}).
On the client side, at least for BFT, there has to be a voting mechanism as part of the framework that selects the correct answers from replicas.

As SMR systems are distributed implementations of a single service or object, the need for some correctness proof is inherent.
In the next section, we will address the relationships between linearizability and SMR systems.

\section{Linearizability and SMR}
\label{sec:linsmr}

\subsection{Historic Observations}
\label{sec:history}

The use of linearizability as a correctness proof of SMR systems, concepts, and approaches is ubiquitous.
This means that it is assumed that an SMR system that is proven to be linearizable has the correct behaviour.

It is not entirely clear to us when researchers started to choose linearizability as a correctness criterion.
Lynch explains in her seminal book~\cite{lynch_distributed_1996} so called \emph{atomic objects} that sometime are also called \emph{linearizable objects}, and connects them to fault-tolerance mechanisms, i.e. to replication.
As already mentioned, Castro chose linearizability as the correctness criterion for his PBFT system~\cite{castro_practical_2001,castro_correctness_1999}.
Multiple times, Castro stressed that the goal was on one hand to ensure that PBFT behaves like a non-distributed system, and on the other hand, that executions are atomic.
Likewise, many papers can be found that choose linearizability as criterion of choice, e.g., Bezerra et al. for an SMR system using data partitioning ~\cite{bezerra_scalable_2014}, Skrzypczak et al. for an SMR system using CRDTs~\cite{skrzypczak_linearizable_2019}, Berger et al. fixing a bug in an optimisation built into PBFT and other replication protocols~\cite{berger_making_2021}.
More or less explicit, some authors argue that hiding the distributed nature of a system is achieved by linearizability, whereas the atomicity of linearizability is not in focus, e.g., Marandi et al.~\cite{marandi_rethinking_2014}, Escobar et al.~\cite{escobar_boosting_2019}.
The given references are a short excerpt of the many works that can be found and could have been mentioned here.

We can summarise that linearizability was chosen because it proves that a distributed object---the replicated state machine---behaves like its non-distributed counterpart.
Sometimes, the atomicity of operations was important, sometimes it was taken as a side effect.

The classic implementation of SMR with an atomic broadcast protocol and sequential execution did not need to focus on the application or object code, as sequential executions are by definition linearizable.
This may have lead to the wrong conclusion that atomic broadcast protocols are already a sufficient condition for linearizability.
This might explain the following citations:
Le et al. claim that SMR guarantees linearizability~\cite{le_dynamic_2016}:
\emph{``State machine replication (SMR) is a well-known technique that guarantees strong consistency (i.e., linearizability) to online services."} is the first sentence of the abstract.
Mendizabal et al. write ~\cite{mendizabal_efficient_2017}: \emph{``Consequently, state machine replication ensures strong consistency (or more
precisely, linearizability)."}
Noguiera et al. set up requirements for their system~\cite{nogueira_elastic_2017}. One of it reads as \emph{``Preserve linearizability: A fundamental property of a RSM\footnote{RSM stands for Replicated State Machine, an alternate term for SMR that is more focussed on the concrete instance than on the concept.} is that it implements strong consistency."}
Pan et al. write~\cite{pan_rabia_2021}: \emph{``SMR ensures linearizability, ..."}
Again, this list could easily be extended.

We already mentioned in Section~\ref{sec:smr} that modern SMR systems might use other protocols instead of atomic broadcasts.
Further, the execution of the application object may not be sequential.
It is obvious that a proof of linearizability has not only to consider the framework but also include the object that is providing the service.
If the object itself, or the state machine respectively, is not linearizable, the underlying framework can try to make it linearizable.
A framework could enforce sequential execution which can achieve linearizability, but unfortunately not for all cases.
In the next section, we will show these cases in which there are apparently object implementations that are not linearizable and the framework is \textbf{not} able to achieve linearizable behaviour.

\subsection{Is it a Match?}
\label{sec:match}

The observations compiled above lead to two research questions:

\begin{itemize}
  \item Is linearizability an appropriate correctness criterion for SMR systems? In other words: \emph{Is it a match?}
  \item Does the concept of SMR automatically lead to linearizable systems, as claimed by some authors?
\end{itemize}

\noindent The answer of the second question follows implicitly from the answer to the first one.

\lstset{language=java,caption={Lock-based multithreading example},label={list:lock},basicstyle=\small}
\begin{lstlisting}

int sharedVar = 1;

void D() {
   int localVar;
   myLock.lock();
     localVar = sharedVar+1; // access
     sharedVar = localVar;
   myLock.unlock();
     localVar = localVar * 2; // compute
   myLock.lock();
     sharedVar = localVar; // access
   myLock.unlock();
}

void E() {
  int localVar; 
  myLock.lock();
    localVar = sharedVar; // access
  myLock.unlock();
  return localVar;
}
\end{lstlisting}

\begin{figure}[ht]
  \center
  \includegraphics[width=.28\textwidth]{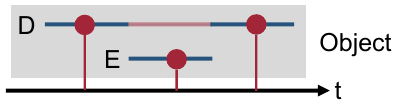}
  \caption{Interrupted invocations with locks become non-linearizable}
  \label{fig:lock}
\end{figure}

Let us first look at SMR research that clearly produces non-linearizable systems.
The first domain that can be found is deterministic multithreading using so called lock-level approaches.
Shared state is protected by locks and a scheduler deterministically assigns locks to concurrent threads, each executing a particular request.
As with deterministic lock assignments the information flow between threads becomes deterministic, threads outside of locks can run concurrently.
This concept is called weak determinism~\cite{olszewski_kendo_2009}.
There are at least four scheduling algorithms that use this technique: PDS and LSA~\cite{basile_active_2006}, MAT~\cite{reiser_consistent_2006} and UDS~\cite{hauck_uds_2016}.\footnote{Kendo is another prominent system that uses a lock-level approach~\cite{olszewski_kendo_2009}. 
Unfortunately, in its original version it is not suitable for application in SMR as it cannot integrate newly arriving requests.
A fix will need a comprehensive modification of its scheduling algorithm.}
All four were implemented and tested in SMR systems.
Although not explicitly mentioned by the authors, their approach in general leads to non-linearizable behaviour.
A request execution may need multiple critical sections that are protected by locks.
All four algorithms will not guarantee that two critical sections of one request execution are not interrupted by another critical section of another execution.
As an example, see Listing~\ref{list:lock} and its possible concurrent execution in Fig.~\ref{fig:lock}.
While the lock is free in the execution of D (light red part of the timeline) the execution of E might get the lock.
The execution of D needs at least two linearization points as intermediate results become visible to the execution of E.
Request executions are no longer atomic and therefore not linearizable.
The example of Listing~\ref{list:lock} should generate squared numbers, e.g., 1, 4, 25 etc.
As the system is not linearizable the read operation E could also read a value 2, if the first invocation of D ist interrupted by E.
A framework could try to convert these non-linearizable objects into linearizable ones by executing their requests sequentially.
This, however, would contradict the effort to have more efficient concurrent executions.

Another interesting aspect was mentioned by Reiser et al.~\cite{reiser_consistent_2006}.
The authors obviously integrated conditional waits into MAT and PDS such that a thread that acquired a lock may release the lock and be blocked until a condition is fulfilled which is signalled by another thread holding the lock.
This is similar to the monitor concept by Hoare~\cite{hoare_monitors_1974}.
Such blocking behaviour that occurs after some effect and before another effect will lead to at least two linearization points and thus is not compatible with linearizability, too.
For conditional waits, a framework is no longer able to turn the non-linearizable object into a linearizable one, because sequential execution will not prevent multiple linearization points within request execution as these are blocked and resumed.
In the blocking interval other requests need to be executed that can observe intermediary results.
If we would prevent these other requests to be executed, the original request can never be woken up.

\begin{figure}[ht]
  \center
  \includegraphics[width=.37\textwidth]{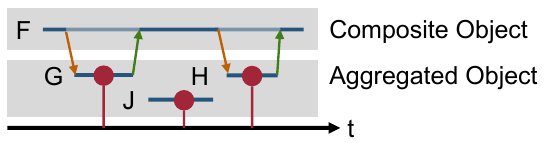}
  \caption{Non-linearizable nested invocations}
  \label{fig.nested}
\end{figure}

Another research area is concerned with nested invocations in SMR systems, e.g.,~\cite{fang_redundant_2004,maassen_efficient_2000,pleisch_replicated_2003,berger_automatic_2022}.
As soon as a replicated service calls other replicated or non-replicated services, the SMR system not only has to deal with the duplications of this nested invocation, but the system will not be linearizable in general.
As an example, consider a request F that needs to invoke requests G and H in an aggregated object.
If another client invokes request J directly in the aggregated object, it might be executed right between G and H as shown in Fig.~\ref{fig.nested}.
This will require two linearization points for F as intermediary results could leak to J.

The only way to achieve linearizability for the calling operation is to ensure that the called object is exclusively used by the caller and nested invocations of the same caller are enforced to execute in the order of the linearization points of the caller.
This can hardly be achieved in general as the aggregated object may have its own framework or may even be out of any control, e.g., as an external service.
In short, as soon as the called service is shared with other services or even other executions of the same caller, the system is no longer linearizable in general.
This makes vertical composition non-linearizable in general, too.

\begin{figure}[ht]
  \center
  \includegraphics[width=.32\textwidth]{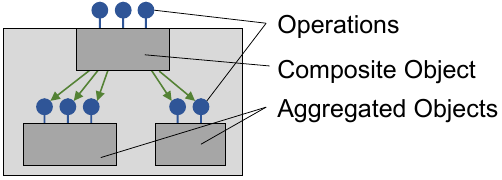}
  \caption{Vertical composition of two objects}
  \label{fig.vertical}
\end{figure}

Vertical composition means that a composite object uses other objects to fullfil its own service.
These other objects are ``vertically" on another level, whereas horizontally composed objects are on the same level (cf. Fig.~\ref{fig.horizontal}).
Even when the aggregated objects used for compositions are linearizable, the composite object using them will not be, at least not in general.
Figure~\ref{fig.vertical} shows an example for vertical composition of two aggregated objects to a new composite.
Note that the composite here, unlike Figure~\ref{fig.horizontal}, contains code that orchestrates the invocations of the aggregated objects.
The operations provided by the composite may also be entirely different from the ones provided by the aggregated objects, whereas horizontal composition just can combine the existing operations of its composed objects.

As a conclusion, we can summarise a first contribution of this paper:

\begin{itemize}
  \item Linearizability cannot be fulfilled as soon as concurrent executions
  \begin{itemize}
    \item exchange data in both directions (linearizability can only have an information flow from previous to later linearization points),
    \item use nested invocations to shared external services,
    \item can block in the middle of their execution with effects before and after the blocking (e.g., conditional waits).
  \end{itemize}
  \item There are SMR systems that are not linearizable and there is no evidence that these systems are somehow incorrect.
  \item As there are counter examples, the statement that SMR guarantees linearizability is wrong, even though published multiple times in well reputed conferences.
\end{itemize}

Of course, linearizability can still be used to prove SMR systems correct.
However, this excludes certain applications that could be replicated using SMR technology.
It might also exclude SMR optimisations, e.g., deterministic multithreading approaches and egalitarian consensus protocols.
Especially interesting is classic linearizability when it comes to data-oriented applications with merely read and write operations, as their semantics typically implies atomic executions.
However, it makes sense to have a different criterion next to linearizability to prove SMR systems correct, in general.

\subsection{Solution}
\label{sec:solution}

\begin{figure}[ht]
  \center
  \includegraphics[width=.31\textwidth]{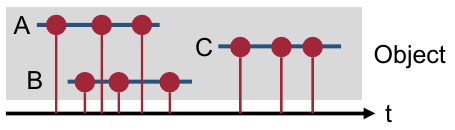}
  \caption{Multi-point linearizability with a multitude of linearization points}
  \label{fig:mplin}
\end{figure}

The counter examples mentioned in Section~\ref{sec:match} all need multiple linearization points during executions.
Thus, a solution could be to introduce a correctness criterion that allows multiple linearization points.
An informal definition could be as follows: Each execution is allowed to have one or more linearization points at which all parts of the execution manifest as atomic events, while all linearziation points are totally ordered in time.
Fig.~\ref{fig:mplin} shows an example of this kind of execution for the scenario of Fig.~\ref{fig:lin}.
Let us call this criterion MP linearizability for multi-point linearizability.

As MP linearizability is some abstract concurrent specification of object behaviour, it is  interval linearizable due to Goubault et al.~\cite{goubault_concurrent_2018}, i.e. all MP-linearizable executions are also interval linearizable:
~\emph{MP Linearizability} $\subset$ \emph{Interval Linearizability}.
We are quite confident that all interval-linearizable executions are also MP linearizable which would make MP linearizability equal to interval linearizability.
Proving this would not add anything to the contributions of this paper so that we omit a proof here.
Instead, we propose to use interval linearizability as the criterion of choice to prove SMR systems correct.

Interval linearizability will allow bidirectional data flow between concurrent, i.e. overlapping, executions.
It will allow nested invocation to shared services, as during the overlap interval arbitrary interaction may occur.
It will even allow conditional waits as these will just extend the overlap interval.
Thus, interval linearizability not only guarantees horizontal composition of objects (cf. local property), but also their vertical composition.
With vertical composition an object uses other objects to implement its own behaviour.
Further, interval linearizability has the required property for distributed systems that lets them appear as a single concurrent object---a requirement that was mentioned by many of the previously cited authors.

Most importantly, interval linearizability preserves some binding to real time.
If subsequent requests do not overlap in execution, the later request must see all effects of the previous one.

As interval linearizability is the property all concurrent specifications of an object have~\cite{goubault_concurrent_2018}, it appears to be the most generic property that could be used.
Any further relaxation would likely give up the above discussed behaviour to be equivalent to a non-distributed system.

Interestingly, in his tutorial about SMR, Schneider did not present any correctness criterion, but gave two properties that clients of an SMR system can rely on~\cite{schneider_implementing_1990}:

\begin{itemize}
  \item[O1] Requests issued by a single client
  to a given state machine $sm$ are processed by $sm$ in the order they were issued.
  \item[O2] If the fact that request $r$ was made to
  a state machine $sm$ by client $c$ could
  have caused a request $r'$ to be made
  by a client $c'$ to $sm$, then $sm$ processes
  $r$ before $r'$.
\end{itemize}

These properties are not sufficient for correctness proofs of SMR systems.
It is however obvious that interval linearizability will fulfill both of them.
In both cases, there is no overlap between the request executions, and the effect of a first request is guaranteed to be observable by the second request.

In summary, a second contribution of this paper is to propose interval linearizability as the correctness criterion for SMR systems as it covers all possible specifications of application objects.
This effectively means that any object is allowed to be replicated with SMR technology, and the SMR framework will guarantee that the replicated (distributed) implementation behaves like the non-replicated one.

Replacing linearizability by interval linearizability forfeits the atomicity property.
As soon as this is a requirement for an application, linearizability remains the property of choice.
Thus, proving SMR systems correct should consider both interval- and classic-linearizable applications, or objects respectively.
Even better and as explained in the next section, an SMR system does not actually need to be proven against one or the other property.

\section{Correctness Proof of SMR Systems}
\label{sec:proof}

\begin{figure}[ht]
  \center
  \includegraphics[width=.33\textwidth]{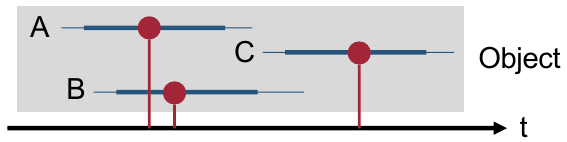}
  \caption{Extended invocation time}
  \label{fig.linext}
\end{figure}

In order to compile how a correctness proof of an SMR system could look like, we remember the distinction between application and framework (cf. Figure~\ref{fig.framework}).
The application cannot be directly accessed by clients.
All access is controlled and managed by the SMR framework.
More precisely, incoming invocation requests first arrive at the framework.
After ordering the requests, the invocation leads to an invocation in the application code.
Let us assume that this application code is proven to be linearizable.
From an outside perspective, the additional framework means that invocations take more time.
The actual invocation event is prior to the invocation event happening when the application code is invoked, as it has to be processed and ordered by the framework.
Likewise, the response event happens after the response event from the application, as the response has to be forward to the client and in BFT case some voting has to take place.
Figure~\ref{fig.linext} shows the scenario of Figure~\ref{fig:lin.1}, but with extended timelines.
The thin parts of the execution timeline show these extensions.

\begin{lemm}
\label{lemm:lext}
  If all execution scenarios of a linearizable object are changed such that all invocation events are prior and all response events are later, then the new executions are linearizable, too. 
\end{lemm}

In order to prove Lemma~\ref{lemm:lext}, the linearizable object will take care of a single linearization point within the inner timeline of the execution (thick lines in Figure~\ref{fig.linext}).
Thus, the extended timelines will also have a single linearization point.
The total order of linearization points is not changed by the extensions of the timelines.
There may be more overlaps between executions, e.g., overlap of request B and C that was not there before.
In theory, that may introduce more degrees of freedom for placing linearization points, but the points can actually only be arranged within the inner timelines.
In summary, the extensions do not change anything on linearization points, which makes the extension linearizable as well.
This leads to the conclusion that as long as the framework is just extending the timelines of a linearizable application object then the entire SMR system is linearizable.

Interestingly, this is also true for interval-linearizable applications.
For interval linearizability only the overlapping intervals of the inner timelines are relevant.
These are not changed, so we can state another lemma that reads:

\begin{lemm}
\label{lemm:ilext}
  If all execution scenarios of an interval-linearizable object are changed such that all invocation events are prior and all response events are later, then the new executions are interval linearizable, too. 
\end{lemm}

According to Lemmata~\ref{lemm:lext} and~\ref{lemm:ilext}, it is irrelevant whether we would like to achieve a linearizable or interval-linearizable SMR system.
As long as the application is linearizable, or interval linearizable respectively, and the framework just leads to timeline extensions the entire system remains linearizable, or interval linearizable respectively.

In the following, we will argue about how we can prove that an SMR framework just extends timelines. First we address the crash-stop and crash-recovery failure models and then the Byzantine model.

\subsection{Crash Failure Models}

We have to prove that the framework just does the mentioned timeline extensions.
However, an SMR system contains multiple distributed replicas.
In the CFT case, only one of them delivers the result to the client.
In this replica, the execution is timeline-extended as discussed above, so that Lemmata~\ref{lemm:lext} and~\ref{lemm:ilext} apply.
However, it is not clear which of the replicas delivers the answer to the client.
In order to be correct, the framework has to ensure that the execution within replicas is deterministic so that it is irrelevant which replica delivers the response---a trivial aspect that is well-known to SMR developers.

\begin{figure}[h]
  \center
  \includegraphics[scale=.55]{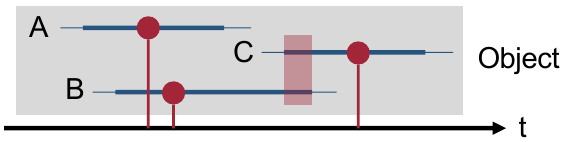}
  \caption{Alternate execution in different replica compared to Figure~\ref{fig.linext}. The shaded area shows an additional overlap interval.}
  \label{fig.linext.repl}
\end{figure}

Note also that the overlaps of a set of invocations may differ in each replica.
Typically, the invocation events of the application object (of the inner timeline) are in the same order in all replicas (atomic broadcast), the response events may be quite differently ordered.
Some invocations may take longer in one replica than in the other.
Figure~\ref{fig.linext.repl}, for example, shows a delayed response event for Request~B such that there is now an additional overlap with the execution of Request~C that was not there in the execution shown in Figure~\ref{fig.linext}.
Theoretically, an extended inner timeline would allow a different arrangement of linearization points in one replica, but not in the other.
However, due to the requirement of a deterministic execution, the order of linearization points will be the same in each replica.

Two linearization points could be swapped in different replicas if their effect is independently created.
This possibility may be used by deterministic multithreading approaches and by egalitarian consensus protocols.
Nevertheless, determinism ensures that the outcome is always equivalent to the execution of any particular replica so that Lemmata~\ref{lemm:lext} and~\ref{lemm:ilext} still apply.
The same concept can be applied to interval linearizability.
If two executions do not depend on each other, it is irrelevant when and how these overlap.
The requirement for a deterministic behaviour will ensure that even then the lemmata can be applied.

Determinism may also affect the execution of the application object compared to a stand-alone usage.
Thus, the SMR framework has to ensure that the deterministic execution does not change the linearizability property.\footnote{For interval linearizability no such proof needs to be made as a deterministic execution of an interval-linearizable object will always remain interval linearizable. This is due to the fact that every concurrent object specification is interval linearizable~\cite{goubault_concurrent_2018}.}

\subsection{Byzantine Failure Model}

In the BFT case, the requirements on determinism are the same as for the CFT case.
However, some of the replicas may now have malicious behaviour and thus expose a completely arbitrary execution pattern which may no longer be linearizable and more importantly not comply to the execution in correct replicas.
Therefore, a proof of a correct SMR framework needs to include a proof of the correct filtering of malicious results within the limits of the failure model---which is again not surprising to SMR developers.
Taking the first result that was equally responded by $f+1$ replicas does the job, if no more than $f$ replicas are malicious.

Special care has to be taken if there are Byzantine faulty clients.
First, such a client could inject different invocations with the same invocation ID into the SMR system.
This case has to be handled so that there is no chance that different replicas execute different requests.
In fact, this is again a requirement for deterministic behaviour in (correct) replicas.

Second, a Byzantine client could incorrectly vote for malicious instead of correct answers by replicas.
From the perspective of such clients the behaviour of the SMR system might be neither linearizable nor interval linearizable.
Similar to Castro~\cite{castro_practical_2001}, such clients have to be excluded from considerations.
Malicious clients only affect their own behaviour and observations but neither the SMR system itself nor other clients.
Thus, excluding malicious clients from correctness proofs is a reasonable procedure.

\subsection{Summary}

The third contribution of this paper is that proving an SMR framework correct should not prove the linearizability or interval linearizability of the entire system. 
Instead, the following steps have to be made for a correctness proof:

\begin{itemize}
  \item Prove that the application object is linearizable. In case interval linearizability is enough, this step can be skipped as every concurrent object specification is interval linearizable.
  \item Prove that the application object is deterministically executed. This includes deterministic multithreading and handling of indeterministic instructions, statements, operations and system calls but excluding input\footnote{We assume that an SMR system does not have any other input than the incoming requests. In case such input is required, it can be modelled as a nested invocation to some special service that deterministically serves the input to replicas, and is subject to its own correctness proof.}.
  In case of a linearizable application object, it has to be proven that the deterministic execution does not compromise linearizability.
  \item Prove that the application object gets deterministic input in form of new requests. 
  This is typically achieved by an atomic broadcast.
  So it has to be proven that the atomic broadcast correctly works under the assumption of the underlying failure model.
  Next to atomic broadcasts, there may be different approaches, e.g. egalitarian partial-order consensus algorithms that have to be proven correct accordingly.
  \item Prove that the framework tolerates faults within the chosen failure model. For the case of BFT, prove that correct clients do a correct voting on responses.
\end{itemize}

As we laid out above, as soon as all aspects are proven correct, the SMR system preserves the linearizability or interval linearizability of the application object in use.
Unfortunately, this is not much less than had to be proven anyway.
However, SMR developers do no longer have to deal with linearizability, and after accepting Section~\ref{sec:match} also not with interval linearizability.
Instead, the steps above ensure that these properties are inherited from the application object that is replicated.

\section{Conclusion}
\label{sec:conclusion}

In this paper, we argued that linearizability is too strict to be used as a correctness criterion for SMR systems.
Nested invocations, i.e. vertical composition, lock-based synchronisation and conditional waits are not allowed given the definition of linearizability, although applications using these techniques can be replicated by SMR systems.
We further argued that interval linearizability, a recent correctness criterion, is the best choice to cover the important aspect that the distributed architecture of an SMR system shows a behaviour that can be experienced from a non-distributed but concurrent application object.
In case it is appropriate, linearizability may still be of advantage, e.g., for storage services with merely read and write operations, as it additionally enforces atomic executions.

We further showed that it is not necessary to prove that the entire SMR system, framework, application object, and their combination, are linearizable or interval linearizable.
Instead we start with a proven linearizable or interval linearizable application object and prove that the framework and its distributed replicas in sum preserve the linearizable or interval-linearizable behaviour of the application object.
The proof of the framework thus comprises the deterministic execution and the deterministic input delivered to the replicated application objects.
Further, the correct voting of results in case of BFT has to be achieved.
Thus, our contribution will shift the focus from linearizability for correctness proofs of SMR systems to correct consensus protocols, correct deterministic executions and correct response voting under the restrictions of a given failure model.
This procedure is the same for linearizable \emph{and} interval-linearizable applications.

Finally, we pointed out that some papers claim that SMR can only be realized under linearizability, or even worse, that the SMR concept itself guarantees linearizability.
We proved these claims to be false.

%
\section*{Acknowledgments}
We thank our fellow researcher for discussing of and commenting on this research, especially the participants of the Reykjavík Summer School on Secure and Reliable Distributed System in 2022.
Thanks go also to the anonymous reviewers who commented on earlier versions of this paper.

\printbibliography

@inproceedings{reiser_consistent_2006,
	series = {{SRDS}},
	title = {Consistent {Replication} of {Multithreaded} {Distributed} {Objects}},
	doi = {10.1109/SRDS.2006.14},
	booktitle = {25th {IEEE} {Symp}. on {Rel}. {Dist}. {Sys}.},
	author = {Reiser, Hans P. and Domaschka, Jörg and Hauck, Franz J. and Kapitza, Rüdiger and Schröder-Preikschat, Wolfgang},
	month = oct,
	year = {2006},
	pages = {257--266},
}

@article{basile_active_2006,
	title = {Active replication of multithreaded applications},
	volume = {17},
	issn = {1558-2183},
	doi = {10.1109/TPDS.2006.56},
	number = {5},
	journal = {IEEE Trans. on Par. and Distr. Sys.},
	author = {Basile, Claudio and Kalbarczyk, Zbigeniew and Iyer, Ravishankar K.},
	month = may,
	year = {2006},
	pages = {448--465},
}

@inproceedings{alchieri_boosting_2018,
	title = {Boosting {State} {Machine} {Replication} with {Concurrent} {Execution}},
	doi = {10.1109/LADC.2018.00018},
	booktitle = {8th {Latin}-{Amer.} {Symp}. on {Dep}. {Comp}.},
	series = {{LADC}},
	author = {Alchieri, Eduardo and Dotti, Fernando and Marandi, Parisa and Mendizabal, Odorico and Pedone, Fernando},
	month = oct,
	year = {2018},
	pages = {77--86},
}

@inproceedings{hauck_uds_2016,
	series = {{SRDS}},
	title = {{UDS}: {A} {Novel} and {Flexible} {Scheduling} {Algorithm} for {Deterministic} {Multithreading}},
	shorttitle = {{UDS}},
	doi = {10.1109/SRDS.2016.030},
	booktitle = {{IEEE} 35th {Symp}. on {Rel}. {Distr}. {Sys}.},
	author = {Hauck, Franz J. and Habiger, Gerhard and Domaschka, Jörg},
	month = sep,
	year = {2016},
	pages = {177--186},
}

@inproceedings{berger_making_2021,
	series = {{SRDS}},
	title = {Making {Reads} in {BFT} {State} {Machine} {Replication} {Fast}, {Linearizable}, and {Live}},
	doi = {10.1109/SRDS53918.2021.00010},
	booktitle = {40th {Int}. {Symp}. on {Rel}. {Distr}. {Sys}.},
	author = {Berger, Christian and Reiser, Hans P. and Bessani, Alysson},
	month = sep,
	year = {2021},
	pages = {1--12},
}

@inproceedings{marandi_rethinking_2014,
	series = {{ICDCS}},
	title = {Rethinking {State}-{Machine} {Replication} for {Parallelism}},
	doi = {10.1109/ICDCS.2014.45},
	booktitle = {{IEEE} 34th {Int}. {Conf}. on {Distr}. {Comp}. {Sys}.},
	author = {Marandi, Parisa Jalili and Bezerra, Carlos Eduardo and Pedone, Fernando},
	month = jun,
	year = {2014},
	pages = {368--377},
}

@inproceedings{mendizabal_efficient_2017,
	address = {Orlando, FL},
	series = {{IPDPS}},
	title = {Efficient and {Deterministic} {Scheduling} for {Parallel} {State} {Machine} {Replication}},
	isbn = {978-1-5386-3914-6},
	url = {https://ieeexplore.ieee.org/document/7967165/},
	doi = {10.1109/IPDPS.2017.29},
	Xlanguage = {en},
	urldate = {2022-03-31},
	booktitle = {{IEEE} {Int}. {Par}. and {Distr}. {Proc}. {Symp}.},
	author = {Mendizabal, Odorico M. and De Moura, Ruda S.T. and Dotti, Fernando Luis and Pedone, Fernando},
	month = may,
	year = {2017},
	pages = {748--757},
}

@article{herlihy_linearizability_1990,
	title = {Linearizability: a correctness condition for concurrent objects},
	volume = {12},
	issn = {0164-0925, 1558-4593},
	shorttitle = {Linearizability},
	url = {https://dl.acm.org/doi/10.1145/78969.78972},
	doi = {10.1145/78969.78972},
	Xlanguage = {en},
	number = {3},
	urldate = {2022-03-31},
	journal = {ACM Trans. on Progr. Lang. and Sys.},
	author = {Herlihy, Maurice P. and Wing, Jeannette M.},
	month = jul,
	year = {1990},
	pages = {463--492},
}

@inproceedings{sela_brief_2021,
	address = {New York, NY, USA},
	series = {{PODC}},
	title = {Brief {Announcement}: {Linearizability}: {A} {Typo}},
	isbn = {978-1-4503-8548-0},
	shorttitle = {Brief {Announcement}},
	url = {https://doi.org/10.1145/3465084.3467944},
	doi = {10.1145/3465084.3467944},
	urldate = {2023-01-20},
	booktitle = {{ACM} {Symp}. on {Princ}. of {Distr}. {Comp}.},
	author = {Sela, Gal and Herlihy, Maurice and Petrank, Erez},
	month = jul,
	year = {2021},
	pages = {561--564},
}

@article{castaneda_unifying_2018,
	title = {Unifying {Concurrent} {Objects} and {Distributed} {Tasks}: {Interval}-{Linearizability}},
	volume = {65},
	issn = {0004-5411},
	shorttitle = {Unifying {Concurrent} {Objects} and {Distributed} {Tasks}},
	url = {https://doi.org/10.1145/3266457},
	doi = {10.1145/3266457},
	number = {6},
	urldate = {2023-01-20},
	journal = {J. of the ACM},
	author = {Castañeda, Armando and Rajsbaum, Sergio and Raynal, Michel},
	month = nov,
	year = {2018},
	pages = {45:1--45:42},
}

@inproceedings{neiger_set-linearizability_1994,
	address = {New York, NY, USA},
	series = {{PODC}},
	title = {Set-linearizability},
	isbn = {978-0-89791-654-7},
	url = {https://doi.org/10.1145/197917.198176},
	doi = {10.1145/197917.198176},
	urldate = {2023-01-20},
	booktitle = {13th Ann. {ACM} Symp. on {Princ.} of Distr. Comp.},
	publisher = {ACM},
	author = {Neiger, Gil},
	month = aug,
	year = {1994},
	pages = {396},
}

@inproceedings{goubault_concurrent_2018,
	address = {Dagstuhl, Germany},
	series = {{LIPIcs}},
	title = {Concurrent {Specifications} {Beyond} {Linearizability}},
	volume = {125},
	isbn = {978-3-95977-098-9},
	url = {http://drops.dagstuhl.de/opus/volltexte/2018/10088},
	doi = {10.4230/LIPIcs.OPODIS.2018.28},
	booktitle = {22nd {Int.} {Conf.} on {Princ.} of {Distr.} {Sys.} ({OPODIS})},
	Xpublisher = {Schloss Dagstuhl–Leibniz-Zentrum fuer Informatik},
	author = {Goubault, Éric and Ledent, Jérémy and Mimram, Samuel},
	Xeditor = {Cao, Jiannong and Ellen, Faith and Rodrigues, Luis and Ferreira, Bernardo},
	year = {2018},
	Xnote = {ISSN: 1868-8969},
	pages = {28:1--28:16},
}

@phdthesis{castro_practical_2001,
	type = {Ph.{D}.},
	title = {Practical {Byzantine} {Fault} {Tolerance}},
	school = {MIT},
	author = {Castro, Miguel},
	month = jan,
	year = {2001},
}

@techreport{castro_correctness_1999,
	type = {Techn. {Rep.}},
	title = {A {Correctness} {Proof} for a {Practical} {Byzantine}-{Fault}-{Tolerant} {Replication} {Algorithm}},
	Xlanguage = {en-US},
	number = {MIT/LCS/TM-590},
	urldate = {2023-09-03},
	institution = {MIT},
	author = {Castro, Miguel and Liskov, Barbara},
	month = jun,
	year = {1999},
}

@article{schneider_implementing_1990,
	title = {Implementing fault-tolerant services using the state machine approach: a tutorial},
	volume = {22},
	issn = {0360-0300},
	shorttitle = {Implementing fault-tolerant services using the state machine approach},
	url = {https://dl.acm.org/doi/10.1145/98163.98167},
	doi = {10.1145/98163.98167},
	number = {4},
	urldate = {2023-10-30},
	journal = {ACM Comp. Surv.},
	author = {Schneider, Fred B.},
	year = {1990},
	pages = {299--319},
}

@article{lamport_time_1978,
	title = {Time, clocks, and the ordering of events in a distributed system},
	volume = {21},
	issn = {0001-0782},
	url = {https://dl.acm.org/doi/10.1145/359545.359563},
	doi = {10.1145/359545.359563},
	number = {7},
	urldate = {2023-10-31},
	journal = {Comm. of the ACM},
	author = {Lamport, Leslie},
	month = jul,
	year = {1978},
	pages = {558--565},
}

@inproceedings{le_dynamic_2016,
	series = {{DSN}},
	title = {Dynamic {Scalable} {State} {Machine} {Replication}},
	url = {https://ieeexplore.ieee.org/abstract/document/7579726},
	doi = {10.1109/DSN.2016.11},
	urldate = {2023-10-31},
	booktitle = {46th {Ann}. {IEEE}/{IFIP} {Int}. {Conf}. on {Dep}. {Sys}. and {Netw}.},
	author = {Le, Long Hoang and Bezerra, Carlos Eduardo and Pedone, Fernando},
	month = jun,
	year = {2016},
	Xnote = {ISSN: 2158-3927},
	pages = {13--24},
}

@inproceedings{bezerra_scalable_2014,
	series = {{DSN}},
	title = {Scalable {State}-{Machine} {Replication}},
	url = {https://ieeexplore.ieee.org/abstract/document/6903591},
	doi = {10.1109/DSN.2014.41},
	urldate = {2023-10-31},
	booktitle = {44th {Ann}. {IEEE}/{IFIP} {Int}. {Conf}. on {Dep}. {Sys}. and {Netw}.},
	author = {Bezerra, Carlos Eduardo and Pedone, Fernando and Van Renesse, Robbert},
	month = jun,
	year = {2014},
	pages = {331--342},
}

@article{nogueira_elastic_2017,
	title = {Elastic {State} {Machine} {Replication}},
	volume = {28},
	issn = {1558-2183},
	url = {https://ieeexplore.ieee.org/abstract/document/7885120},
	doi = {10.1109/TPDS.2017.2686383},
	number = {9},
	urldate = {2023-10-31},
	journal = {IEEE Trans. on Par. and Distr. Sys.},
	author = {Nogueira, Andre and Casimiro, Antonio and Bessani, Alysson},
	month = sep,
	year = {2017},
	pages = {2486--2499},
}

@inproceedings{pan_rabia_2021,
	address = {New York, NY, USA},
	series = {{SOSP}},
	title = {Rabia: {Simplifying} {State}-{Machine} {Replication} {Through} {Randomization}},
	isbn = {978-1-4503-8709-5},
	shorttitle = {Rabia},
	url = {https://dl.acm.org/doi/10.1145/3477132.3483582},
	doi = {10.1145/3477132.3483582},
	urldate = {2023-10-31},
	booktitle = {{ACM} 28th {Symp}. on {Oper}. {Sys}. {Princ}.},
	author = {Pan, Haochen and Tuglu, Jesse and Zhou, Neo and Wang, Tianshu and Shen, Yicheng and Zheng, Xiong and Tassarotti, Joseph and Tseng, Lewis and Palmieri, Roberto},
	year = {2021},
	pages = {472--487},
}

@book{lynch_distributed_1996,
	address = {San Francisco, CA, USA},
	title = {Distributed {Algorithms}},
	isbn = {978-0-08-050470-4},
	publisher = {Morgan Kaufmann},
	author = {Lynch, Nancy A.},
	year = {1996},
}

@incollection{merideth_selected_2010,
	Xaddress = {Berlin, Heidelberg},
	series = {LNCS},
	number = 5959,
	title = {Selected {Results} from the {Latest} {Decade} of {Quorum} {Systems} {Research}},
	isbn = {978-3-642-11294-2},
	url = {https://doi.org/10.1007/978-3-642-11294-2_10},
	language = {en},
	urldate = {2024-01-29},
	booktitle = {Replication: {Theory} and {Practice}},
	publisher = {Springer},
	author = {Merideth, Michael G. and Reiter, Michael K.},
	Xeditor = {Charron-Bost, Bernadette and Pedone, Fernando and Schiper, André},
	year = {2010},
	doi = {10.1007/978-3-642-11294-2_10},
	pages = {185--206},
}

@inproceedings{kapritsos_all_2012,
	title = {All about {Eve}: {Execute}-{Verify} {Replication} for {Multi}-{Core} {Servers}},
	isbn = {978-1-931971-96-6},
	shorttitle = {All about {Eve}},
	Xlanguage = {en},
	urldate = {2024-02-07},
	booktitle = {10th {USENIX} {Symp}. on {Oper}. {Sys}. {Des}. and {Impl}.},
	series = {{OSDI}},
	author = {Kapritsos, Manos and Wang, Yang and Quema, Vivien and Clement, Allen and Alvisi, Lorenzo and Dahlin, Mike},
	year = {2012},
	pages = {237--250},
}

@inproceedings{kapitza_storyboard_2010,
	title = {Storyboard: {Optimistic} {Deterministic} {Multithreading}},
	shorttitle = {Storyboard},
	Xlanguage = {en},
	urldate = {2024-02-07},
	booktitle = {6th {Worksh}. on {Hot} {Topics} in {Sys}. {Dep}.},
	series = {{HotDep}},
	publisher = {USENIX},
	author = {Kapitza, Rüdiger and Schunter, Matthias and Cachin, Christian and Stengel, Klaus and Distler, Tobias},
	year = {2010},
}

@inproceedings{eischer_egalitarian_2021,
	title = {Egalitarian {Byzantine} {Fault} {Tolerance}},
	url = {https://ieeexplore.ieee.org/abstract/document/9667692?casa_token=og0LTw9W2KEAAAAA:nzff4xQXAR-IzA4el9zdbifGg9yWMztWWoVpbnza_DInTTyrnxRZgmSTB94Ely7Uk4N_twMqe3QY},
	doi = {10.1109/PRDC53464.2021.00019},
	urldate = {2024-02-07},
	booktitle = {{IEEE} 26th {Pacific} {Rim} {Int}. {Symp}. on {Dep}. {Comp}.},
	author = {Eischer, Michael and Distler, Tobias},
	series = {{PRDC}},
	month = dec,
	year = {2021},
	pages = {1--10},
}

@inproceedings{moraru_there_2013,
	address = {New York, NY, USA},
	series = {{SOSP}},
	title = {There is more consensus in {Egalitarian} parliaments},
	isbn = {978-1-4503-2388-8},
	url = {https://dl.acm.org/doi/10.1145/2517349.2517350},
	doi = {10.1145/2517349.2517350},
	urldate = {2024-02-07},
	booktitle = {24th {ACM} {Symp}. on {Oper}. {Sys}. {Princ}.},
	author = {Moraru, Iulian and Andersen, David G. and Kaminsky, Michael},
	month = nov,
	year = {2013},
	pages = {358--372},
}

@inproceedings{olszewski_kendo_2009,
	address = {New York, NY, USA},
	series = {{ASPLOS}},
	title = {Kendo: efficient deterministic multithreading in software},
	isbn = {978-1-60558-406-5},
	shorttitle = {Kendo},
	url = {https://dl.acm.org/doi/10.1145/1508244.1508256},
	doi = {10.1145/1508244.1508256},
	urldate = {2024-02-07},
	booktitle = {14th {Int}. {Conf}. on {Arch}. {Supp}. for {Progr}. {Lang}. and {Oper}. {Sys}.},
	publisher = {ACM},
	author = {Olszewski, Marek and Ansel, Jason and Amarasinghe, Saman},
	year = {2009},
	pages = {97--108},
}

@inproceedings{herlihy_axioms_1987,
	address = {New York, NY, USA},
	series = {{POPL}},
	title = {Axioms for concurrent objects},
	isbn = {978-0-89791-215-0},
	doi = {10.1145/41625.41627},
	urldate = {2024-02-07},
	booktitle = {14th {ACM} {SIGACT}-{SIGPLAN} {Symp}. on {Princ}. of {Progr}. {Lang}.},
	author = {Herlihy, Maurice P. and Wing, Jeannette M.},
	year = {1987},
	pages = {13--26},
}

@inproceedings{skrzypczak_linearizable_2019,
	address = {New York, NY, USA},
	series = {{PODC}},
	title = {Linearizable {State} {Machine} {Replication} of {State}-{Based} {CRDTs} without {Logs}},
	isbn = {978-1-4503-6217-7},
	url = {https://dl.acm.org/doi/10.1145/3293611.3331568},
	doi = {10.1145/3293611.3331568},
	urldate = {2024-02-07},
	booktitle = {{ACM} {Symp}. on {Princ}. of {Distr}. {Comp}.},
	author = {Skrzypczak, Jan and Schintke, Florian and Schütt, Thorsten},
	month = jul,
	year = {2019},
	pages = {455--457},
}

@inproceedings{escobar_boosting_2019,
	address = {New York, NY, USA},
	title = {Boosting concurrency in {Parallel} {State} {Machine} {Replication}},
	isbn = {978-1-4503-7009-7},
	url = {https://dl.acm.org/doi/10.1145/3361525.3361549},
	doi = {10.1145/3361525.3361549},
	urldate = {2024-02-07},
	booktitle = {20th {Int}. {Middlew}. {Conf}.},
	publisher = {ACM},
	author = {Escobar, Ian Aragon and Alchieri, Eduardo and Dotti, Fernando Luís and Pedone, Fernando},
	year = {2019},
	pages = {228--240},
}

@inproceedings{fang_redundant_2004,
	series = {{EEE}},
	title = {A redundant nested invocation suppression mechanism for active replication fault-tolerant {Web} service},
	url = {https://ieeexplore.ieee.org/abstract/document/1287282},
	doi = {10.1109/EEE.2004.1287282},
	urldate = {2024-02-07},
	booktitle = {{IEEE} {Int}. {Conf}. on e-{Techn}., e-{Comm}. \& e-{Service}},
	author = {Fang, Chen-Liang and Liang, Deron and Chen, Chyouhwa and Lin, PuSan},
	month = mar,
	year = {2004},
	pages = {9--16},
}

@inproceedings{maassen_efficient_2000,
	address = {New York, NY, USA},
	series = {{JAVA}},
	title = {Efficient replicated method invocation in {Java}},
	isbn = {978-1-58113-288-5},
	url = {https://dl.acm.org/doi/10.1145/337449.337486},
	doi = {10.1145/337449.337486},
	urldate = {2024-02-07},
	booktitle = {{ACM} {Conf}. on {Java} {Grande}},
	author = {Maassen, Jason and Kielmann, Thilo and Bal, Henri E.},
	month = jun,
	year = {2000},
	pages = {88--96},
}

@techreport{pleisch_replicated_2003,
	type = {Techn. {Rep.}},
	title = {Replicated {Invocation}},
	institution = {EPFL Sci. Pub.},
	author = {Pleisch, Stefan and Kupsys, Arnas and Schiper, André},
	year = {2003},
}

@inproceedings{berger_automatic_2022,
	series = {{EDCC}},
	title = {Automatic {Integration} of {BFT} {State}-{Machine} {Replication} into {IoT} {Systems}},
	url = {https://ieeexplore.ieee.org/abstract/document/9933240},
	doi = {10.1109/EDCC57035.2022.00013},
	urldate = {2024-02-07},
	booktitle = {18th {Eur}. {Dep}. {Comp}. {Conf}.},
	author = {Berger, Christian and Reiser, Hans P. and Hauck, Franz J. and Held, Florian and Domaschka, Jörg},
	month = sep,
	year = {2022},
	pages = {1--8},
}

@article{hoare_monitors_1974,
	title = {Monitors: an operating system structuring concept},
	volume = {17},
	issn = {0001-0782},
	shorttitle = {Monitors},
	url = {https://dl.acm.org/doi/10.1145/355620.361161},
	doi = {10.1145/355620.361161},
	number = {10},
	urldate = {2024-02-14},
	journal = {Comm. of the ACM},
	author = {Hoare, C. Antony R.},
	year = {1974},
	pages = {549--557},
}

@article{papadimitriou_serializability_1979,
	title = {The serializability of concurrent database updates},
	volume = {26},
	issn = {0004-5411},
	url = {https://dl.acm.org/doi/10.1145/322154.322158},
	doi = {10.1145/322154.322158},
	number = {4},
	urldate = {2024-02-14},
	journal = {J. of the ACM},
	author = {Papadimitriou, Christos H.},
	year = {1979},
	pages = {631--653},
}
    
    

\end{document}